\documentclass[pre,reprint,showpacs,onecolumn]{revtex4-2}
\usepackage{amssymb,amsmath}
\usepackage{natbib}
\usepackage{graphicx}
\usepackage{color}
\usepackage{bm}
\usepackage{soul}

	 \renewcommand{\vec}[1]{\mathbf{#1}}

\graphicspath{{./}{figs/}}

\begin{document}

\title{Effective stochastic model for chaos in the Fermi-Pasta-Ulam-Tsingou chain}

\author{Tomer Goldfriend}
\email{tomergf@gmail.com}
\affiliation{Department of Physics of Complex Systems, Weizmann Institute of Science, Rehovot 76100, Israel}

\date{\today}

\begin{abstract}

Understanding the interplay between different wave excitations, such as phonons and localized solitons, is crucial for developing coarse-grained descriptions of many-body, near-integrable systems.  We treat the Fermi-Pasta-Ulam-Tsingou (FPUT) non-linear chain and show numerically that at short timescales, relevant to the  largest Lyapunov exponent, it can be modeled as a random perturbation of its integrable approximation--- the Toda chain. At low energies, the  separation between two trajectories that start at close proximity is dictated by the interaction between few soliton modes and an intrinsic, apparent bath representing a background of many radiative modes. It is sufficient to consider only one randomly perturbed Toda soliton-like mode to explain the   power-law profiles reported in previous works, describing how the Lyapunov exponent of large FPUT chains decreases with the energy density of the system.   
\end{abstract}


\maketitle

\section{Introduction}
\label{sec:intro}

Recently, there is a new interest in many-body classical, and quantum integrable systems, following  experimental advances in solid-state physics~\cite{Kinoshita_etal2004} and new analytical technics to treat exactly solvable models~\cite{Calabrese_etal2016}. This led to a generalized thermodynamic and hydrodynamic formulations for integrable models~\cite{Rigol_etal2007,Vidmar&Rigol2016,CastroAlvaredo_etal2016,Bertini_etal2016,Spohn2019,Doyon2019}, where a macroscopic set of conserved quantities dictates in- and out-of-equilibrium phenomena.  When integrability is slightly broken, two natural questions arise: (A) Do the weakly broken conserved quantities continue to govern the dynamics? and (B) if yes, do all quantities contribute equally?

The answer to the second question depends on the model and the physical process in question, and actually, this question is relevant for integrable systems as well. For example, one can ask whether solitary waves are crucial for the anomalous heat transport in near-integrable (or integrable) non-linear chains~\cite{Li_etal2010}. The answer to question (A) above is typically positive: in near-integrable systems there is a generic separation between Lyapunov and diffusion timescales, which implies that quasi-conserved quantities dictate the slow dynamics of such systems~\cite{Lam&Kurchan2014,Goldfriend&Kurchan2020}. Recently, based on this observation, coarse-grained models were shown to describe the slow thermalization of isolated 1D classical chains, using numerical simulations~\cite{Goldfriend&Kurchan2019}, and of 1D quantum gases, using the formalism of generalized hydrodynamics~\cite{Bastianello2020}.

In the current paper we address the shortest timescales for which an effective, coarse-grained model can be relevant to describe the dynamics of quasi-integrable systems--- the Lyapunov times. We focus on the  Fermi-Pasta-Ulam-Tsingou (FPUT) chain and show that replacing the fast variables with noise can {\it quantitatively} capture the typical timescales for chaos in the system, which are much shorter than equilibration times. Using numerical simulations we can explore the conserved modes of the underlying integrable model--- the Toda chain, and study the different roles played by radiative and soliton-like modes.

Lyapunov exponents measure chaos as the exponential rate at which the distance between two initial conditions in close proximity diverges in time. The maximal Lyapunov exponent is defined as 
\begin{equation}
\lambda\equiv \lim_{t\rightarrow \infty}\lim_{|\vec{u}(0)|\rightarrow 0} \frac{1}{2t}\ln \frac{|\vec{u}(t)|^2}{|\vec{u}(0)|^2},
\label{eq:Lypdef}
\end{equation}
where $\vec{u}(t)=\vec{x}^{(1)}(t)-\vec{x}^{(2)}(t)$ is the separation between two trajectories in phase-space. For integrable systems, all the Lyapunov exponents vanish, as the flow (in action-angle variables) is laminar. In the presence of small integrability breaking perturbation, the Kolmogorov-Arnold-Moser (KAM) theorem states that phase-space breaks into a mixed state of regular regions and unstable chaotic ones. However, in the case of many degrees of freedom, for any practical reason,  the KAM theorem is irrelevant, since the transition at which the non-chaotic islands have a non-zero measure is expected to be exponentially small with the system's size~\cite{Falcioni_etal1991,Pettini_etal2005}.

One way to describe the chaotic behavior of many-body, quasi-integrable systems beyond the KAM regime is the one suggested in Ref.~\cite{Lam&Kurchan2014}: integrability breaking is replaced by a time-dependent random drive. The resulting dynamics has a positive Lyapunov exponent which scales as $\sigma^{1/3}$, where $\sigma$ being the variance of the external noise. This result qualitatively explains the aforementioned typical separation between Lyapunov and equilibration times ($\sim\sigma$ for the randomly driven system) in deterministic, many-body quasi-integrable system. Well known examples of this phenomenon are the Solar System, which has a Lyapunov time of $\sim 5$~Myrs but stability times larger than $5$~Gyrs~\cite{Laskar2008}, or the FPUT chain whose equilibration times can be larger by more than four orders of magnitudes than their Lyapunov times~\citep{Benettin_etal2013}  (see also Refs.~\cite{Mithun_etal2019,Danieli_etal2019} for more recent examples).

The main result of the current Paper is to provide a {\it quantitative} correspondence between the deterministic, complex dynamics of the quasi-integrable FPUT chain and the dynamics of a randomly perturbed integrable Toda chain. In particular, we show, with the aid of numerical simulations, that the Lyapunov separation in the FPUT model is governed by the dynamics of several soliton-like modes of the Toda chain perturbed by an effective bath of radiative (linear) modes. This result is somewhat surprising, since solitons are considered as a signature of integrability.  We find that a minimal model of a single  soliton subjected to a random drive captures how the Lyapunov exponent decreases with the energy, or equivalently, with integrability breaking, in large FPUT chains.

Developing such minimal  models is crucial for understanding the complex dynamics of many-body, quasi-integrable systems. For example, previously it was shown how an effective model for the dynamics of Mercury  yields the short Lyapunov exponent of the Solar System, and how the stochastic features of this model allow to explore orbital instabilities at long times~\cite{Batygin_etal2015}, or short rare events of destabilization~\cite{Woillez&Bouchet}.

The rest of the Paper is organized as follows: In Sec.~\ref{sec:models} we introduce the FPUT and the Toda chains, and in Sec.~\ref{sec:eff} we present our minimal stochastic model.  The main results are given in Sec.~\ref{sec:results}, where we show an agreement between this effective model and the deterministic FPUT system. Finally, we discuss the results in Sec.~\ref{sec:diss}.

\section{The FPUT and the Toda models}
\label{sec:models}

Consider a one-dimensional chain of size $N+1$ with fixed ends, whose Hamiltonian has the general form 
\begin{equation}
\mathcal{H}_{\rm chain}=\frac{1}{2}\sum^N_{n=1} p_n^2 + \sum^N_{n=0} V (q_{n+1}-q_{n}),\quad q_0=q_{N+1}=0,
\label{eq:H_General}
\end{equation}
where $(q_n,p_n)$ are the coordinate and momentum of the $n$th bead, whose unit mass is set to 1, and $V(r)$ is the nearest-neighbor potential. We focus on the chaotic dynamics of the $\alpha+\beta$ FPUT chain, whose potential is: 
\begin{equation}
V_{\rm FPUT}(r)=\frac{1}{2}r^2+\frac{\alpha}{3}r^3+\frac{\beta}{4}r^4,
\label{eq:V_FPUT}
\end{equation}
where we set the time units to 1, and hereafter we also fix $\alpha=1$. The dynamics of the chain thus depends on its size $N$, its energy density $\epsilon\equiv\mathcal{H}_{\rm FPUT}/N$, and the parameter $\beta$. At low energies, the FPUT chain can be considered as a small perturbation to the linear chain, which has the potential $V_{\rm L}(r)=\frac{1}{2}r^2$. The corresponding Hamiltonian $\mathcal{H}_{\rm L}$ is integrable, and its distance from the non-integrable FPUT Hamiltonian can be quantified as  $|V_{\rm L}-V_{\rm FPUT}|\sim \alpha r^3\sim  \alpha\epsilon^{3/2}$. A closer integrable approximation to $\mathcal{H}_{\rm FPUT}$ is the nonlinear, integrable Toda chain, whose potential reads:
\begin{equation}
V_{\rm Toda}(r)= V_0 \left(e^{A r}-1- A r \right).
\label{eq:V_Toda}
\end{equation}
Hereafter we choose the parameters $V_0=(2\alpha)^{-2}$ and $A=2\alpha$, which implies $|V_{\rm Toda}-V_{\rm FPUT}|\sim \beta r^4\sim  \beta \epsilon^2$.

Throughout the paper we work with different coordinate systems which we now define. The first system is the usual phase-space,  designated with $\vec{x}\equiv (\vec{q},\vec{p})$. The second coordinate system, the {\it normal modes} $\vec{X}\equiv (\vec{Q},\vec{P})$, corresponds to the linear chain, and is  defined as   
\begin{align}
\begin{split}
Q_k &\equiv \omega_k \sqrt{\frac{2}{N+1}}\sum_{n=1}^N q_n \sin\left(\frac{\pi k n}{N+1}\right), \\
P_k	&\equiv  \sqrt{\frac{2}{N+1}}\sum_{n=1}^N p_n \sin\left(\frac{\pi k n}{N+1}\right),
\label{eq:FT}
\end{split}
\end{align}
where $\omega_k\equiv2\sin\left(\frac{\pi k}{2(N+1)}\right)$. Eq.~\eqref{eq:FT} is simply the Fourier space, where we normalize the Fourier variable of positions with $\omega_k$. In these coordinates the Hamiltonian of the linear chain is  $\mathcal{H}_{\rm L}=\sum_k \frac{1}{2}(P_k^2+Q_k^2)\equiv \sum_k E_k$, and  one can clearly see that the normal modes' energies $E_k$ are proportional to the action variables of the linear chain. To these we add the normal modes' angles
\begin{equation}
\phi_k\equiv \arctan\left(\frac{P_k}{Q_k}\right),
\end{equation}
which form our third coordinate system $(\phi_k,E_k)$. Next we present our last set of coordinates, which is based on the Toda model.

\subsection{Toda modes}

We construct a specific set of conserved quantities for the Toda Hamiltonian, which we term hereafter as the {\it Toda modes} $\{J_k\}^N_{k=1}$, based on gaps between eigenvalues of Lax matrices. The properties of this set of quantities, as well as their relevance to the FPUT dynamics,  were originally introduced by Ferguson et. al.~\cite{Ferguson_etal1982}, and were discussed more recently in Ref.~\cite{Goldfriend&Kurchan2019}, where it was shown how the slow evolution of $J_k$ governs the equilibration of the FPUT chain. 

Before defining the conserved quantities, let us indicate their key properties  which are crucial for the theoretical arguments in the next Sections: (A) The $J_k$-s are proportional to the canonical action variables of the Toda chain, namely, the number of excited modes corresponds to the dimension of the invariant tori. (B) In the limit of low energies, the Toda modes are proportional to the normal modes of the linear chain, $E_k$. (C) Excitation of low Toda mode numbers $k\sim 1$ corresponds to localized waves, whereas high mode numbers correspond to Fourier modes. Hereafter we refer to the former and the latter as {\it soliton} (see comment at the end of this Section)  and {\it radiative}  modes of the periodic Toda chain, respectively.  In more detail, the Toda spectrum contains two parts: one part of localized excitations $1\leq k \leq k_s$, and a second radiative part, $k_s< k\leq N$, in which a higher mode number $k$ is related to a shorter wavelength.

We now provide the mathematical definition for the Toda modes, a complete discussion and a demonstration on how this definition is related to properties (A)--(C) above are given in Refs.~\cite{Ferguson_etal1982,Goldfriend&Kurchan2019}. We start with extending our fixed ends chain of size $N+1$ to a chain of size $N'=2N+2$, which is periodic and asymmetric: $x'_n=x_n$ and $x'_{N+1+n}=-x_{N+1-n}$ for $n=1\dots N+1$. Then, we define the Flaschka variables~\cite{Flaschka1974} $a_n = \frac{1}{2} e^{\alpha(q'_n-q'_{n-1})}$, $b_n = \alpha p'_{n-1}$, and construct two symmetric Lax matrices, $\vec{L}^+$ and $\vec{L}^-$, of size $N'\times N'$:
\begin{equation}
\vec{L}^{\pm}=
\begin{pmatrix}
b_1     &a_1     &       &        &        &\pm a_{N'}  \\
a_1     &b_2     &a_2    &        &        &        \\	
        &\ddots  &\ddots &\ddots  &        &        \\	
        &        &\ddots &\ddots  &\ddots  &        \\
        &        &       &a_{N'-2}&b_{N'-1}&a_{N'-1}\\
\pm a_{N'}  &        &       &        &a_{N'-1}&b_{N'}  \\
\end{pmatrix},
\label{eq:Lpm}
\end{equation}
where the unoccupied entries are zero. The eigenvalues of these matrices do not vary in time under the Toda dynamics~\cite{Flaschka1974}, and gaps between (part of) these $2N'$ eigenvalues define our $N$ Toda modes as: 
\begin{equation}
\{J_k\}^{N}_{k=1}=\{\lambda_{2k}-\lambda_{2k+1}\}^{k=N}_{k=1},
\label{eq:Jk}
\end{equation} 
where $\{\lambda_n\}^{2N'}_{n=1}$ is the {\it combined spectrum} of both matrices, given in a decreasing order $-\lambda_1\leq\dots\leq -\lambda_{N'-1}\leq 0\leq 0\leq \lambda_{N'-1}\leq\dots\leq  \lambda_1$. Note that for the definition of $J_k$ in Eq.~\eqref{eq:Jk} we take only the positive part of the spectrum, which is asymmetric, and assume that $N=(N'-2)/2$ is odd.  Soliton-like excitations correspond to modes that involve eigenvalues $\lambda_n$ which are larger than 1~\cite{Ferguson_etal1982}. Therefore, the separation between the soliton and radiative parts of the spectrum, at $k=k_s$, is defined as the largest mode number $k$ such that $\lambda_{2k+1}>1$.

We can formally add the angle variables $\{\Psi_k\}^N_{k=1}$ of the Toda model to define our fourth coordinate system,  $(\vec{\Psi},\vec{J})$.  The quasi-periodic motion of the integrable Toda chain is thus given by: $\dot{\vec{J}}=0$, $\dot{\vec{\Psi}}=\vec{\Omega}$, where the Toda frequencies $\vec{\Omega}$ are functions of the initial values of Toda modes. The explicit definition of the canonical angle variables of the Toda chain~\cite{Henrici&Kappeler2008} is not necessary for our work as we demonstrate the results numerically. For example, motion along the angle coordinates is given by evolving the Toda dynamics  in the usual phase-space $\vec{x}$. In Sec.~\ref{sec:ResultB} we discuss how we can numerically move in the space spanned by the Toda modes $J_k$.

Finally, let us comment that the terminology of soliton excitations used throughout the text is slightly inaccurate, since an exact, single traveling soliton exists only at an infinite Toda chain. However, as shown in Ref.~\cite{Ferguson_etal1982}, truncated solitons, or superposition of solitons result in low Toda modes excitations of a finite chain. 

\section{Effective random model}
\label{sec:eff}

The Lyapunov exponent of an integrable system slightly perturbed with noise was explored in detail in Ref.~\cite{Lam&Kurchan2014}: The idea is to study the separation between pairs of trajectories which feel the same realization of the noise. For weak noise, the resulting Lyapunov time-scales were found to be much smaller than the diffusion time-scales induced by the noise, and the separation of trajectories is roughly confined to the invariant tori. It was suggested that replacing integrability breaking interactions with a time-dependent random drive can faithfully describe the deterministic dynamics of many-body quasi-integrable systems. In the current work we show that this is indeed the case for the FPUT chain.

Before introducing our specific model of a perturbed Toda chain, let us recap the known results for a generic, randomly perturbed integrable Hamiltonian~\cite{Lam&Kurchan2014,Goldfriend&Kurchan2020}. It is sufficient to consider a one-dimensional,  dynamical random system  
\begin{align}
\begin{split}
\dot{I} &= g(I,\theta)\eta(t),\\
\dot{\theta} &= \omega(I),
\end{split}
\label{eq:effgen}
\end{align}
where $(I,\theta)$ are action-angle variables, $g$ is some function, and $\eta$ is a Gaussian white noise of zero mean and variance~$\sigma$. Eq.~\eqref{eq:effgen} can result from a randomly perturbed  integrable Hamiltonian $\mathcal{H}=\mathcal{H}_{\rm int}(I)+\mathcal{H}_p(I,\theta)\eta(t)$ (the random drive on $\theta$ is omitted from the dynamical equation since it does not effect the dynamics in the weak noise limit). 
The Lyapunov exponent of this system was found to be~\cite{Lam&Kurchan2014}
\begin{equation}
\lambda \sim \left(\frac{d\omega}{d I} \right )^{2/3} \left(\langle g^2\rangle_\theta \sigma  \right )^{1/3},
\label{eq:Lyp}
\end{equation}
where $\langle \cdot \rangle_{\theta}$ is an average over the angle variable. As mentioned above, the Lyapunov separation occurs mostly along the angle direction, thus, $I$ is assumed to be constant in Eq.~\eqref{eq:Lyp}. 

There are three important ingredients which bring forth the result in Eq.~\eqref{eq:Lyp}: ({\it i}\,) The action variable is subjected to a {\it multiplicative, $\theta$-dependent}, random drive. ({\it ii}\,) The integrable part is {\it nonlinear}, i.e., $\frac{d\omega}{dI}\neq 0$. ({\it iii}\,) There is a {\it separation of timescales} $\tau_c\ll\omega^{-1}\ll \lambda^{-1}$, were $\tau_c$ is the correlation time of the noise (which goes to $0$ in the white noise limit)~\footnote{The separation of $\omega^{-1}$, which gives the typical time around the invariant tori, is not expected to affect the scaling of the Lyapunov exponent.}.

Let us now go back to the deterministic FPUT model, whose dynamics in terms of the Toda variables can be formally written as:
\begin{align}
\begin{split}
\dot{J}_k &= F_k(\vec{J},\vec{\Psi}), \\
\dot{\Psi}_k &=\Omega_k(\vec{J})+\varpi_k(\vec{J},\vec{\Psi}),
\end{split}
\label{eq:eff0_0}
\end{align}
where $\vec{F}$ and $\vec{\varpi}$ are the integrability breaking perturbation terms. We are seeking to write a model in view of Eq.~\eqref{eq:effgen}, from which one can obtain the Lyapunov exponent of the FPUT chain. The central claim of the current paper, which is verified in detail in Sec.~\ref{sec:results}, is the following: the Lyapunov instability of the FPUT chain originates from an effective, random time-dependent perturbation on the Toda solitons. Namely, considering only $F_k$ for small  $1\leq k\leq k_s$ and its stochastic modeling, is enough to quantify chaos in the FPUT chain.  

The rationale behind our approach is that for large systems, it is plausible that the complex integrability-breaking perturbation acts effectively as noise. The seemingly random drive on one part of the system, in our case the solitons, has very short correlation time. This correlation time is intrinsic to the system, emerging from interplay with other parts of the system--- the radiative modes in our case. This picture is somewhat analogous to features of chaos in the Solar System. For example, Ref.~\cite{Hayes_etal2010} showed that while the complete system of Sun + 8 planets is chaotic, a system of Sun + 4 inner planets is not. The conclusion was that it is the interplay between the outer and terrestrial (inner) planets that gives rise to chaos in the Solar System.

In what follows, it is further assumed that the interaction between solitons does not play a role in our model for chaos, and thus we can take a very simplified model where only one Toda soliton feels a random drive. In addition, we drop the angular perturbation $\vec{\varpi}$ in Eq.~\eqref{eq:eff0_0} since it gives higher order corrections to the Lyapunov exponent~\cite{Lam&Kurchan2014,Woillez&Bouchet2017}. An effective stochastic model for the FPUT dynamics can thus be written in the Toda coordinates $(\vec{\Psi},\vec{J})$ as
\begin{align}
\begin{split}
\dot{J}_k &= \delta_{k,1} f(\vec{J},\vec{\Psi})\eta(t), \\
\dot{\Psi}_k &=\Omega_k(\vec{J}),
\end{split}
\label{eq:eff0}
\end{align}
where $f$ is some unknown phase-space function discussed below, and $\eta(t)$ is a Gaussian white noise with
\begin{equation}
\langle \eta(t) \rangle=0, \quad \langle \eta(t)\eta(t') \rangle=\sigma \delta(t-t').
\label{eq:Wnoise}
\end{equation}

If the {\it deterministic} expression for $F_1(\vec{J},\vec{\Psi})\equiv\dot{J}_1(\vec{J},\vec{\Psi})$ under the FPUT dynamics was given, then,  together with  the assumption that this function has sufficiently short correlation time, one could obtain its stochastic asymptotic $f(\vec{J},\vec{\Psi})\eta(t)$ written in Eq.~\eqref{eq:eff0}. This is called homogenization in the theory of stochastic averaging~\cite{Gardiner}. We do not have the analytic form of  $F_1(\vec{J},\vec{\Psi})$, however, we can show numerically that indeed it has a short correlation time, as well as evaluate  its magnitude in the white noise limit. In addition, the specific form of $f(\vec{J},\vec{\Psi})$ is irrelevant for obtaining the scaling of the Lyapunov exponent with the energy density $\epsilon$. All of these statements are examined in the next Section. 

 Finally, let us remark that the lack of energy conservation in our effective model, induced by the external drive, does not affect the results and conclusions concerning the Lyapunov exponent. This is because the change in energy, coming from the diffusion of $J_1$, occurs on timescales that are much larger than the Lyapunov time.

Next, we analyze the simple stochastic model in Eq.~\eqref{eq:eff0}. It is written in the Toda space $(\vec{\Psi},\vec{J}\,)$, to which we do not have analytical, or complete numerical access: the Toda modes $J_k$ are complicated functions of the phase-space $\vec{x}$, and the Toda frequencies $\Omega_k$ are not known. Yet, we can explore numerically the essential properties of this model and show that the resulting Lyapunov exponent agrees with the deterministic one. In particular, the model can be used to estimate the scaling of $\lambda$ with the energy density $\epsilon$.

\section{Results}
\label{sec:results}

Below we show  how the stochastic model in Eq.~\eqref{eq:eff0} can explain the Lyapunov   instability in FPUT chains. We present three different results: (A) we study the  instantaneous change of the Toda modes under the FPUT dynamics,  $F_k$ in Eq.~\eqref{eq:eff0_0}, and show that their autocorrelation for small $k$ (solitons) resembles a random signal with short correlation times, whereas for large $k$ (radiative modes) the signal is correlated for longer times.  (B) We combine theory, namely  Eq.~\eqref{eq:Lyp}, and numerical simulations to find how the Lyapunov exponent of the randomly perturbed Toda chain scales with the energy density $\epsilon$, and show that it agrees with the scaling found in a previous work~\cite{Benettin_etal2018}. (C) We preform a direct simulation for an effective model--- a Toda chain with an additional random drive acting {\em only} on the first Toda mode, $J_1$. We find an agreement between the Lyapunov exponent of this stochastic system, and the one of the deterministic  FPUT chain. For parts (A) and (C) we treat a chain with $N=63$ points, whereas for part (B) we study a larger chain with $N=1023$.

The Lyapunov exponent is examined for initial conditions drawn from an equilibrium state of the normal modes, namely for each $k$, $E_k$ is taken from an exponential distribution with mean $\epsilon$, and $\phi_k$ from a uniform distribution in the interval $[0 ,2\pi]$. The numerical simulations we present below contain different elements. One numerical procedure is integrating the  Hamiltonian dynamics of the FPUT and Toda chains. This is done  using a 4-th order St{\"o}rmer-Verlet method according to Yoshida scheme~\cite{Yoshida1990}, with time-steps ranging from 0.01 to 0.1. We have verified that decreasing time-step results in increasing accuracy of energy conservation (and conservation of Toda modes  for the Toda dynamics) over time. Other, less standard parts of the computation concern the Toda modes, and are explained in the following subsections \ref{sec:ResultA}-\ref{sec:ResultB}, whereas the discussion on numerical errors appears in  Appendix~\ref{sec:Num}.

\subsection{Autocorrelations of Toda modes in the FPUT dynamics}
\label{sec:ResultA}

\begin{figure} 
\centerline{\resizebox{0.45\textwidth}{!}{\includegraphics{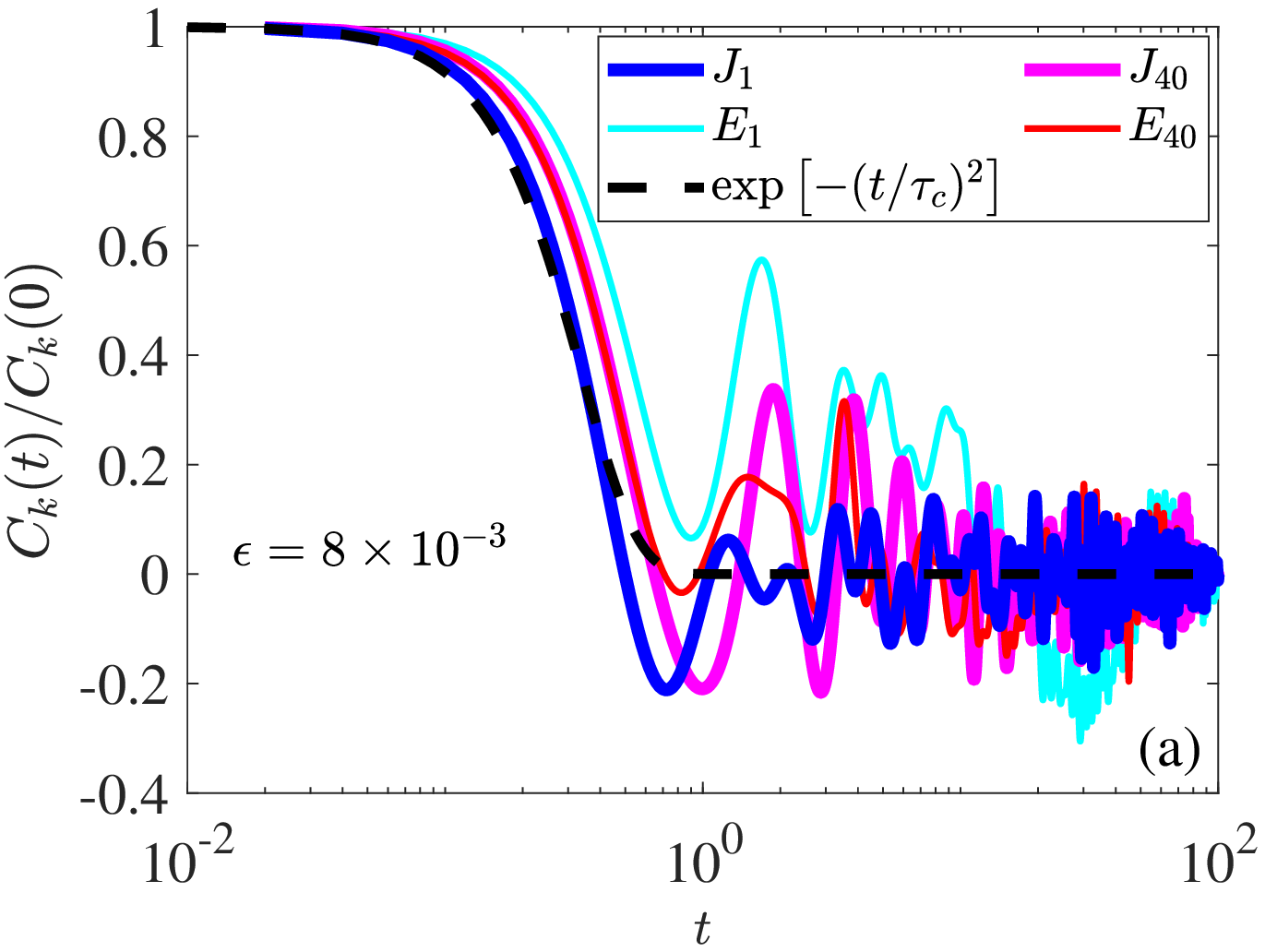}}
\hspace{1cm}
\resizebox{0.45\textwidth}{!}{\includegraphics{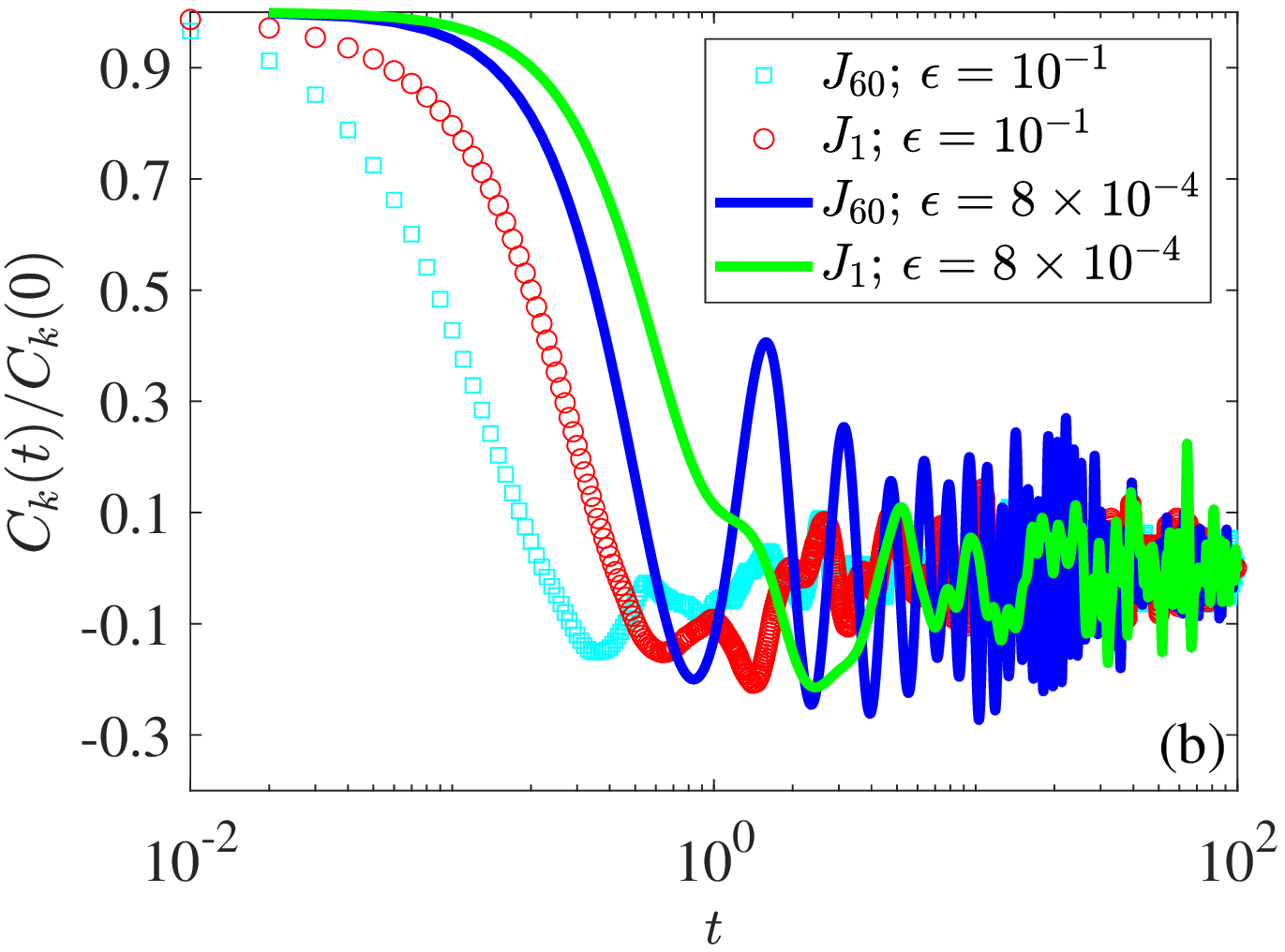}}}
\caption[]{ Autocorrelation for the time-derivative of the Toda modes (Eq.~\eqref{eq:Cdef}) and the normal modes ((Eq.~\eqref{eq:Cdef} with replacing $J_k\rightarrow E_k$) under the FPUT dynamics, for different mode numbers $k$ and energy density $\epsilon$. (a) After a short correlation time, $\tau_c$, the FPUT perturbation induces an incoherent signal on the soliton mode $J_1$ of the Toda chain. The perturbation on higher Toda modes ($J_{40}$), or on the normal modes ($E_1$, $E_{40}$), shows a slower decay. (b) Comparison between the autocorrelation related to the Toda modes $k=1$ and $k=60$, at high  and low energies. The plots are for a single initial condition at a given energy density $\epsilon$.}
\label{fig:corr}
\end{figure}

Before treating the stochastic model directly, we verify its applicability by testing whether under the deterministic FPUT dynamics the soliton Toda modes are effectively driven by a random force. We integrate a chain of size $N=63$ with the FPUT dynamics, calculate the  instantaneous change of the Toda modes' spectrum, and analyze its autocorrelation:
\begin{equation}
C_k(t)=\frac{1}{T_f}\int_0^{T_f} \left[\dot{J}_k(t'+t)-\bar{\dot{J}}_k\right]\left[\dot{J}_k(t')-\bar{\dot{J}}_k\right] dt',
\label{eq:Cdef}
\end{equation}
where $\bar{\dot{J}}_k\equiv \frac{1}{T_f}\int_0^{T_f} \dot{J}_k(t')dt'$, and $T_f$ is the final time of integration.
In Fig.~\ref{fig:corr} we plot $C_k(t)$ for various mode numbers $k$, and energy density $\epsilon$. Computation of the Toda modes $J_k$ includes finding the eigenvalues of the Lax matrices $\vec{L}^{\pm}$ in Eq.~\eqref{eq:Lpm}. We note that this  numerical operation can be facilitated by transforming the almost tridiagonal Lax matrices to an exact pentadiagonal ones, see Ref.~\cite{Ferguson1980} or Appendix in Ref.~\cite{Goldfriend&Kurchan2019}. The time-derivative of $J_k$ is obtained by evolving the system with the FPUT dynamics for an infinitesimal time-step $dt\sim 10^{-4}-10^{-6}$.

As shown in Fig.~\ref{fig:corr}(a), the correlation for the  Toda mode $k=1$ falls rapidly after some typical time $\tau_c$. On the other hand, the  instantaneous change of higher Toda modes stays correlated for longer times. Hence, it is plausible to think of the radiative modes as acting effectively as a random drive on the solitons.  The  Figure also presents  autocorrelations for the normal (Fourier) modes, namely, Eq.~\eqref{eq:Cdef} with replacing $J_k\rightarrow E_k$. As expected for these linear modes, lower mode numbers are correlated for longer times.

Recall that the lower the energy density $\epsilon$, the closer is the system to the linear chain, and hence, the closer are the Toda modes to the linear modes. Therefore, we expect that at very small $\epsilon$ we have only few soliton excitations, which effectively feel a random drive, whereas at high energies even high mode numbers show an uncorrelated behavior.  This is demonstrated in Fig.~\ref{fig:corr}(b), where we compare between the behavior of the first and the 60th Toda modes at small energy $\epsilon=8\times 10^{-4}$, and present how at high energy density of $\epsilon=0.1$ both modes evolve incoherently in a similar fashion.

In summary, the results presented in Fig.~\ref{fig:corr} support the idea presented by the effective model in Eq.~\eqref{eq:eff0}.

\subsection{Scaling of $\lambda(\epsilon)$}
\label{sec:ResultC}

Next, we use our model to obtain the scaling of $\lambda$ as a function of the energy density $\epsilon$. The Lyapunov exponent of Eq.~\eqref{eq:eff0} can be found by extending the result for one degree of freedom in Eq.~\eqref{eq:Lyp} to the case of many-degrees of freedom. This was done in Ref.~\cite{Lam&Kurchan2014}. In general, we shall have   
\begin{equation}
\lambda \sim \left| \frac{\partial \vec{\Omega}}{\partial J_1}\right|^{2/3}\left(\langle f^2\rangle_{\vec{\Psi}} \sigma\right)^{1/3}.
\label{eq:Lypeps}
\end{equation} 
Below we extract numerically the two factors on the right-hand-side of Eq.~\eqref{eq:Lypeps} and find their dependence on the energy density $\epsilon$. The first factor is a property of the integrable Toda chain, whereas the second one depends on the integrability breaking by the FPUT model. We focus on a chain of size $N=1023$.

We start with $\tilde{\sigma}\equiv\langle f^2\rangle_{\vec{\Psi}} \sigma$, which is essentially the variance of the apparent random drive acting on the Toda soliton $J_1$. One way to obtain this quantity is to examine how the mean-square-displacement of $J_1$ in the FPUT dynamics grows in time: The idea is to fit the deterministic time evolution of $J_1$ to a  dynamics generated by a series of kicks, whose magnitude scales as the square-root of the time between kicks.  Given the evolution of $J_1(t)$ at time-interval $[0,T_f]$, we can coarse-grain its temporal changes over time windows of size $\Delta T\gg \tau_c$,  
\begin{equation}
\bar{J}^{\Delta T}_1(n)\equiv J_1\left(n\Delta T\right)-J_1\left(n\Delta T-\Delta T\right) ,\qquad n=1,\dots \frac{T_f}{\Delta T}.
\label{eq:JDeltaT}
\end{equation}
If we claim that the first Toda mode is effectively driven by a white  noise, then the variance of $\bar{J}^{\Delta T}$, over $T_f/\Delta T$ points, shall scale linearly with $\Delta T$--- the slope being   $\tilde{\sigma}$.  Note that for this procedure the specific form of the function $f(\vec{J},\vec{\Psi})$ is irrelevant, as long as $T_f$ is larger than the unperturbed period on the tori.

\begin{figure}
\centerline{\resizebox{0.45\textwidth}{!}{\includegraphics{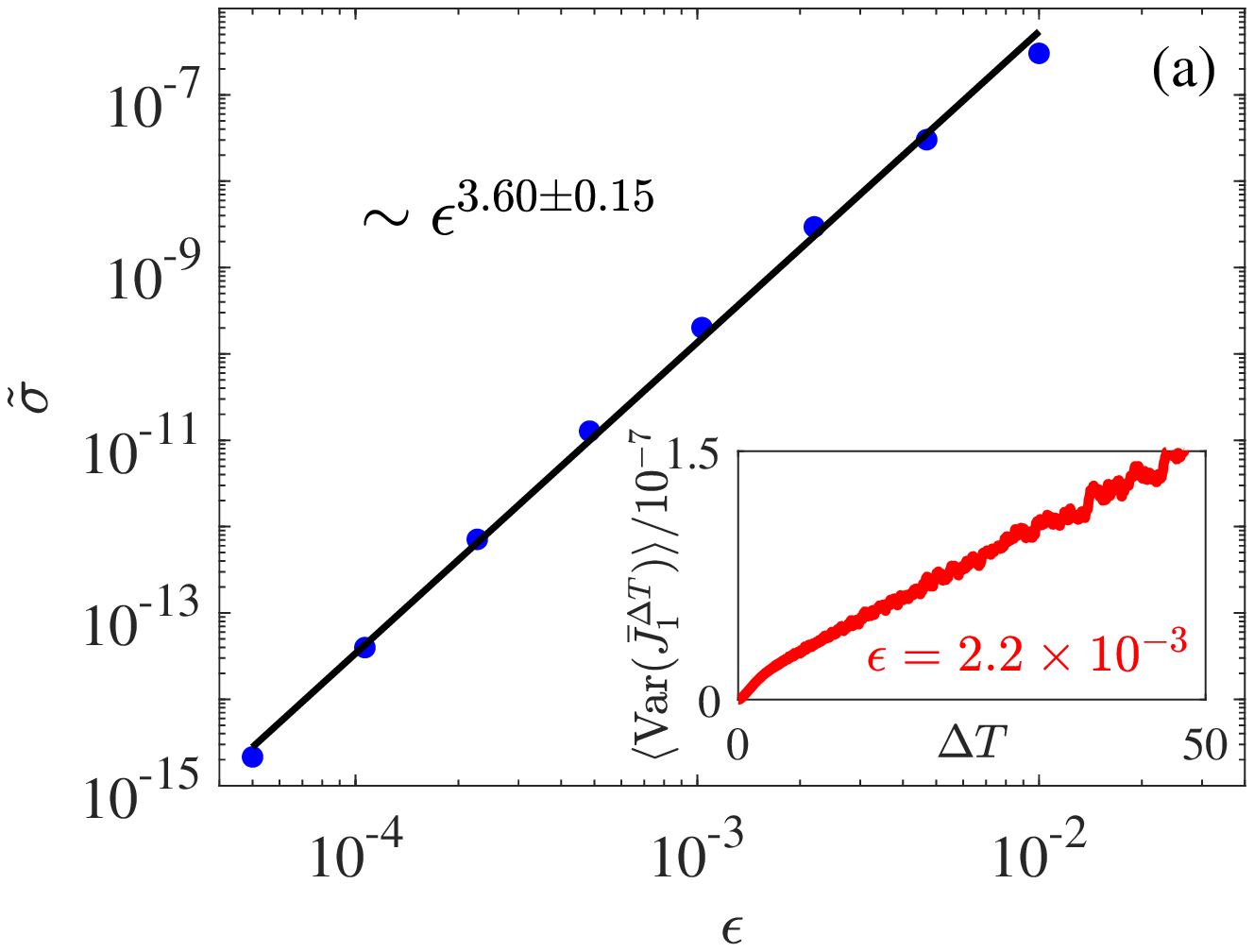}}
\hspace{1cm}
\resizebox{0.45\textwidth}{!}{\includegraphics{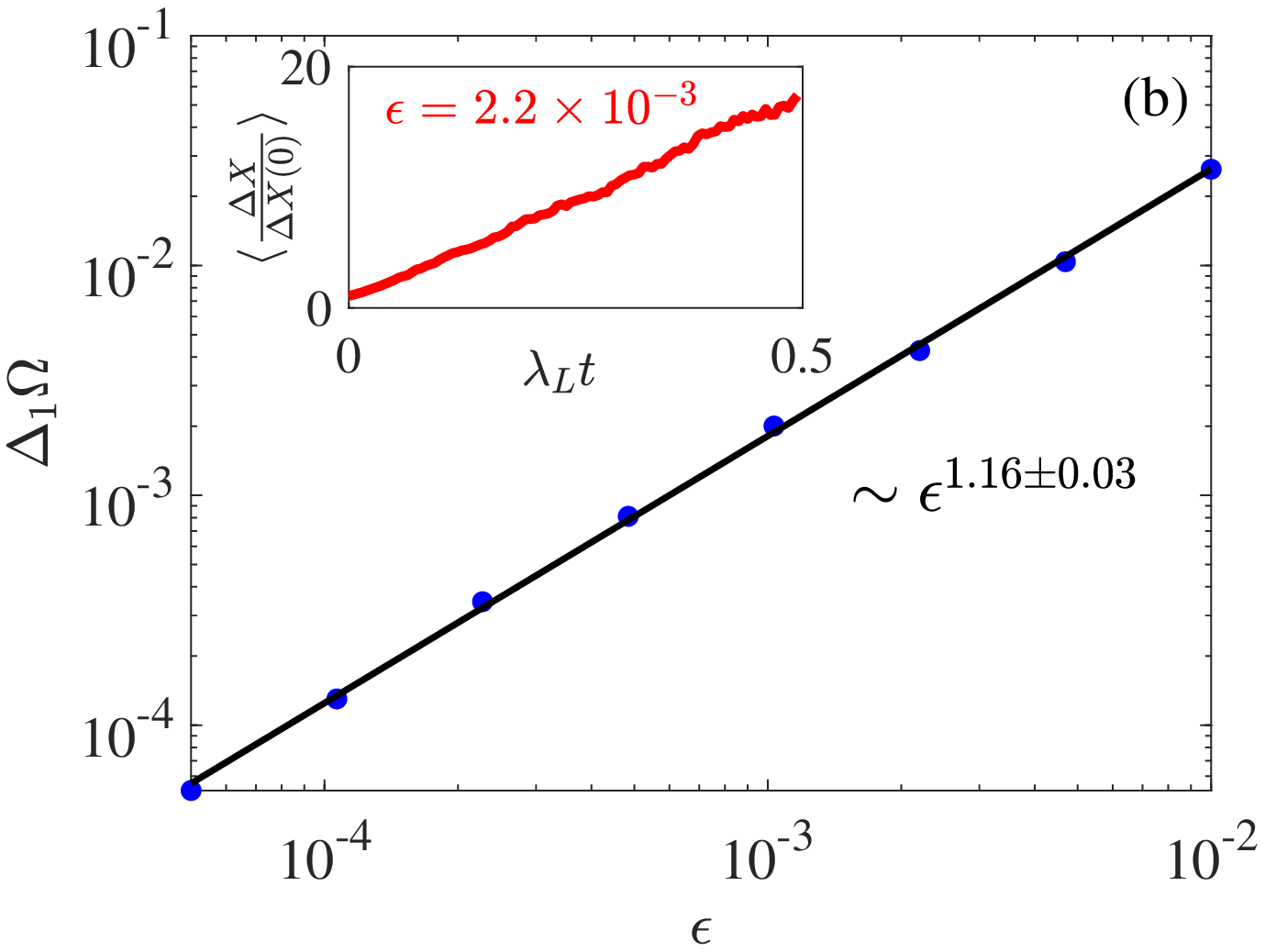}}}
\caption[]{Scaling of the quantities which govern the Lyapunov exponent according to the theoretical prediction in Eq.~\eqref{eq:Lypeps}. (a)~The magnitude of the effective random drive acting on $J_1$. Each point in the main plot is obtained by fitting the linear growth of the coarse-grained evolution of $J_1$ under the FPUT dynamics, an example given in the inset. (b) The angular velocity of the integrable Toda chain, associated  with $J_1$ and affecting the Lyapunov exponent value. Each point in the plot is obtained by examining the linear separation between two initial conditions which are initially separated along the $J_1$ direction, and evolve with the Toda Hamiltonian. An example of this linear separation is given in the inset. The curves in both insets are averaged over 100 initial conditions.}
\label{fig:scaling}
\end{figure}

In the inset of Fig.~\ref{fig:scaling}(a) we show that indeed the variance of $\bar{J}^{\Delta T}$ has a clear linear growth at large enough $\Delta T$, where we average over initial conditions.  Therefore, for each energy density we can fit a linear line to the averaged $\langle {\rm Var}\left(\bar{J}^{\Delta T}_1\right) \rangle$, whose slope gives us $\tilde{\sigma}(\epsilon)$. One can see from Fig.~\ref{fig:scaling}(a)  that $\tilde{\sigma}(\epsilon)$ follows a power-law $\tilde{\sigma}(\epsilon)\sim \epsilon^{\gamma_1}$, and we find $\gamma_1={3.6\pm 0.15}$.

Next, we need to extract the power-law for $\left| \frac{\partial \vec{\Omega}}{\partial J_1}\right|$,  which refers to the relative velocity between points on two adjacent tori, $J_1$ and $J_1+\Delta J_1$ within  the integrable Toda dynamics.  To this end,  we take two initial conditions in phase space, which differ only by their first Toda mode, $(\Delta\vec{J}(0),\Delta\vec{\Omega}(0))=(\Delta J_1,0,\dots,0)$ (see discussion in Sec.~\ref{sec:ResultB} below on how such a configuration can be obtained). When integrated with the {\it Toda} dynamics we get, on average, two trajectories that get further away from each other {\it linearly} in time, with a constant velocity  given by $\Delta_1\Omega\equiv |\frac{\partial \vec{\Omega}}{\partial J_1}dJ_1|$~\cite{Goldfriend&Kurchan2020}. In the inset of Fig.~\ref{fig:scaling}(b) we provide an example of this linear separation. The profile and the power-law fit of $\Delta_1\Omega(\epsilon)\sim \epsilon^{\gamma_2}$ is plotted in Fig.~\ref{fig:scaling}(b), where we find $\gamma_2=1.16\pm 0.03$. The last ingredient we need for evaluating $\lambda(\epsilon)$ is the scaling of $J_1(\epsilon)$, which can be computed directly by averaging the values of the first Toda mode for different initial conditions, for a given $\epsilon$. We find (the result not shown here) $J_1(\epsilon)\sim \epsilon^{\gamma_3}$, with $\gamma_3=0.63$ (the certainty for the fitted exponent $\gamma_3$ is smaller than $0.01$).

We can now collect all the different scaling laws and estimate
\begin{equation}
\lambda \sim \left(\frac{\Delta_1 \Omega(\epsilon)}{J_1(\epsilon)}\right)^{2/3}\tilde{\sigma}(\epsilon)^{1/3}\sim \epsilon^{\frac{2(\gamma_2-\gamma_3)+\gamma_1}{3}}\sim \epsilon^{1.55 \pm 0.07}.
\end{equation} 
This result is in excellent agreement with the one reported by Benettin et. al.~\cite{Benettin_etal2018}, where it was found, by calculating the Lyapunov exponent in the standard way, that $\lambda \sim \epsilon^{1.57}$ for $N=1023$.

\subsubsection{Other FPUT-like chains}

Following Ref.~\cite{Benettin_etal2018} we can study other quasi-Toda models, the $\beta_T$ and the $\gamma_T$ chains, whose potentials are:
$$
V_{\beta_T}(r)=\frac{1}{2}r^2+\frac{\alpha}{3}r^3+\frac{2\alpha^2}{3}\frac{1}{4}r^4,\quad \text{and}\quad 
V_{\gamma_T}(r)=V_{\beta_T}(r)+\frac{\alpha^3}{3}\frac{1}{5}r^5,
$$
respectively. These models are closer to the Toda chain than the usual $\alpha+\beta$ FPUT,  $|V_{\rm Toda}-V_{\beta_T}| \sim   \epsilon^{5/2}$ and $|V_{\rm Toda}-V_{\gamma_T}|\sim  \epsilon^{3}$. According to the theory presented here, the Lyapunov exponent of these systems is dictated by the apparent random drive induced on the Toda soliton mode,  $\tilde{\sigma}(\epsilon)$. 
    
We simulate the $\beta_T$ and $\gamma_T$ models for short timescales, up to $T_f=50-100$, and repeat the numerical procedure to find $\tilde{\sigma}(\epsilon)$ (as given in Fig.~\ref{fig:scaling}(a) for the $\alpha+\beta$ model). We find the exponents $\tilde{\sigma}_{\beta_T}(\epsilon)\sim \epsilon^{4.53\pm 0.25}$ and $\tilde{\sigma}_{\gamma_T}(\epsilon)\sim \epsilon^{5.53\pm 0.09}$. Plugging these exponents to the expression in Eq.~\eqref{eq:Lypeps}, we find $\lambda_{\beta_T}\sim\epsilon^{1.86\pm 0.10}$ and $\lambda_{\gamma_T}\sim\epsilon^{2.19\pm 0.05}$. These scaling laws are with an excellent agreement with Ref.~\cite{Benettin_etal2018}, who found the exponents  $1.9$ and $2.2$ for the $\beta_T$ and $\gamma_T$ models respectively (these reported values are the typical ones for different chain sizes $N$, and not for the specific $N=1023$).

The results for the quasi-Toda models indicate the following hierarchy: decreasing the perturbation magnitude of the Toda {\em Hamiltonian} as $ \epsilon^2\rightarrow\epsilon^{5/2}\rightarrow\epsilon^3$ for the $(\alpha+\beta)\rightarrow\beta_T\rightarrow\gamma_T$ models respectively, corresponds to perturbing the Toda first {\em soliton mode} with the decreasing strength $\epsilon^a\rightarrow\epsilon^{a+1}\rightarrow\epsilon^{a+2}$, with $a\sim 3.5$ for $N=1023$. This interesting result means that the apparent random force on $J_1$ must depend on $J_1$ itself. This fact is crucial for the longer relaxation time of the system; See discussion in Sec.~\ref{sec:diss} below.

We now move to a direct numerical simulation of the effective stochastic model.

\subsection{Weakly, and randomly kicked Toda chain}
\label{sec:ResultB}

The random model in Eqs.~\eqref{eq:eff0}--\eqref{eq:Wnoise} cannot be directly simulated since the function $f(\vec{J},\vec{\Psi})$, which defines integrability breaking interactions in the Toda space, is unattainable due to the complex transformation $(\vec{p},\vec{q})\rightarrow (\vec{J},\vec{\Psi})$. However, the Lyapunov exponent of the stochastic model is insensitive to the specific form of this function, since: (a) any dependence on the angle variables $\Psi_k$ is averaged out, as the Lyapunov time is assumed to be larger than the unperturbed period on the tori, and (b) any dependence on the Toda modes  $J_k$ enters as a constant factor since the diffusion time for $J_1$ is larger than the Lyapunov time. Namely, an equivalent model, which shall yield the same Lyapunov exponent as the model in Eqs.~\eqref{eq:eff0}--\eqref{eq:Wnoise}, reads
\begin{align}
\begin{split}
\dot{J}_k &= \delta_{k,1} \tilde{f}(\vec{\Psi})\tilde{\eta}(t), \\
\dot{\Psi}_k &=\Omega_k(\vec{J}),
\end{split}
\label{eq:eff1}
\end{align}
where $\tilde{\eta}(t)$ is a Gaussian white noise with variance $\tilde{\sigma}\equiv\langle f^2\rangle_{\vec{\Psi}} \sigma$ and $\tilde{f}$ is only a $\Psi-$dependent function which averages to 1 over the invariant Toda tori, $\langle \tilde{f}^2\rangle_{\vec{\Psi}}=1$. Hereafter we take the function 
\begin{equation}
\tilde{f}\equiv \frac{1}{\mathcal{N}} \sum^N_{k=1} \cos\phi_k,
\label{eq:mult}
\end{equation}
where $\phi_k$ are the angles of the normal modes (see Sec.~\ref{sec:models}), and $\mathcal{N}$ is a normalization factor such that $\langle \tilde{f}^2\rangle_\Psi=1$. Again, although the specific relation between $\phi_k$ and $\Psi_k$ is unknown, the normalization factor can be evaluated numerically, by averaging $\sum^N_{k=1} \cos\phi_k$ over the Toda dynamics.

We now describe how Eq.~\eqref{eq:eff1} can be numerically integrated. The way to simulate this random model is  to treat the noise as a series of random kicks,  whose magnitude  are taken from a Gaussian distribution of zero mean
\begin{equation}
\tilde{\eta}(t)=\sum^{\infty}_{i=1} r_i \sqrt{\tau_{\rm kick}} \delta(t-i \tau_{\rm kick}), \quad r_i\sim \mathcal{N}(0,\tilde{\sigma}).
\end{equation}
Therefore, the dynamics of the effective system involves free Toda evolution and instantaneous random kicks. Considering the state of the system just before the $i$-th kick results in the following map
\begin{align}
\begin{split}
J^{(i+1)}_k &= \delta_{k,1} \tilde{f}(\vec{x}^{(i)})r_{i}\sqrt{\tau_{\rm kick}}, \\
\Psi^{(i+1)}_k &= \Psi^{(i)}_k+\Omega_k(\vec{J}^{(i+1)})\tau_{\rm kick},
\label{eq:map}
\end{split}
\end{align}
where the function $\tilde{f}$ is given in Eq.~\eqref{eq:mult}.

As emphasized in Sec.~\ref{sec:eff}, the random effective model is relevant for timescales which are of order of the Lyapunov time $t_{\rm Lyp}=\lambda^{-1}$. In particular, the model prescribes a random motion with zero mean to $J_1$, which is a {\it positive} variable. In theory,   $J_1$ shall stay positive  since we know it does not vary significantly during the Lyapunov regime. However, this point might be crucial for the numerical simulations, where the white noise is represented by random kicks of finite size. One way to overcome this problem is to add a drift term to the equation of $\dot{J}_1$ which is large enough to guarantee that the value of $J_1$ will not become negative, but small enough not to affect the Lyapunov exponent. This technical issue, which becomes irrelevant as $\epsilon$ and $\tau_{\rm kicks}$ (and thus the kicks' magnitude) decreases, is described in detail in Appendix~\ref{sec:drift}.

The variance of the random kicks, $\tilde{\sigma}$, {\it is chosen according to the effective random drive acting on $J_1$ in the non-stochastic, deterministic FPUT dynamics}. This is taken from the curve of $\tilde{\sigma}(\epsilon)$ obtained in the previous Sec.~\ref{sec:ResultC}, see Eq.~\eqref{eq:JDeltaT}. For the rate of kicking we choose 200-600 kicks per Lyapunov time, this guarantees the time-scale separation between noise and Lyapunov times, see discussion after Eq.~\eqref{eq:Lyp}. We have checked that taking different kicking rates, $200\tau_{\rm kicks}=t_{\rm Lyp}$ or $600\tau_{\rm kicks}=t_{\rm Lyp}$, does not change the result.

The free Toda evolution between kicks is preformed numerically in the usual phase-space $\vec{x}$. The kicks are also done in this space, by finding a translation $\vec{dx}$ which results in $(\vec{dJ},\vec{d\Omega})=(dJ_1,0,0,\dots,0)$. We solve this inverse problem by using projection matrices: consider a matrix $\vec{B}$ whose columns are given by a set of $\ell$ vectors of size $2N$, $\vec{B}=\left(\vec{v}_1,\dots,\vec{v}_{\ell}\right)$, then the matrix
\begin{equation}
\vec{P}^{\perp}_{\left\{\vec{v}_k\right\}} \equiv \vec{I}_{2N\times 2N}-\vec{B}\left(\vec{B}^T \vec{B}\right)^{-1}\vec{B}^T,
\label{eq:proj}
\end{equation}
projects vectors of size $2N$ into the space perpendicular to all $\vec{v}_k$. The matrix $\vec{I}_{2N\times 2N}$ is the identity matrix. 

To induce a motion along the $J_1$ direction we first compute vectors normal to the surface of constant $J_k$, $\hat{\vec{n}}_k\equiv \partial_\vec{x} J_k/|\partial_\vec{x} J_k|$. The spatial derivatives of $J_k$  can be obtained numerically by computing the eigenvectors of the $\vec{L}^{\pm}$ matrices~\cite{Goldfriend&Kurchan2019}. Since $J_k$ are not the canonical action variables of the Toda chain we have that the unit vectors $\hat{\vec{n}}_k$ are not orthogonal to each other; but, they {\it are orthogonal} to the direction of the angles $\Psi_k$. We can thus find the $\vec{d\Psi}$ direction using the projection matrix $\vec{P}^{\perp}_{\{\vec{\hat{n}_k}\}}$ and multiplying it with $N$ random vectors $\vec{u}_k$ to define $\vec{\hat{m}}_k\equiv\vec{P}^{\perp}_{\{\vec{\hat{n}_k}\}}\vec{u}_k/|\vec{P}^{\perp}_{\{\vec{\hat{n}_k}\}}\vec{u}_k|$. The set of $N$ unit vectors $\hat{\vec{m}}_k$ spans the space of small variations in the direction of the angles $\vec{\Psi}$. As a final step, we use the projection matrix for the space perpendicular to all $\left\{\vec{\hat{n}}_k,\vec{\hat{m}}_k\right\}$, except $\vec{\hat{n}}_1$--- this gives us a direction in phase-space along which small translation $\vec{dx}$ results in changing $J_1$ alone.

\begin{figure}
\centerline{\resizebox{0.45\textwidth}{!}{\includegraphics{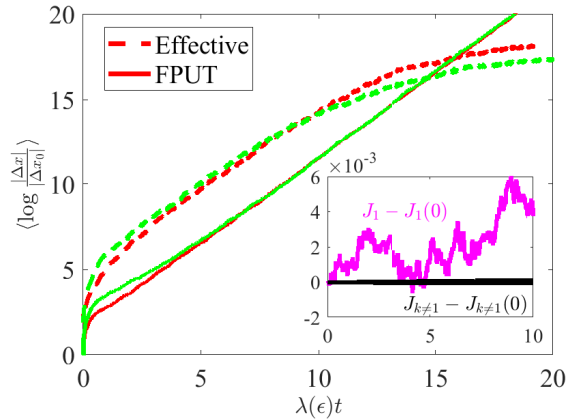}}}
\caption[]{Comparison between Lyapunov separation in the FPUT dynamics (solid lines) and in the effective stochastic model (dashed lines) for chain size $N=63$. We show examples for two different energy densities, $\epsilon=10^{-3}$ (red/dark line) and $\epsilon= 2.6\times 10^{-4}$ (green/bright line). The quenched average for both models is over $\sim 100$ initial conditions, where for the stochastic model we also average over noise realizations. The time axis is rescaled with the Lyapunov exponent for each of the two examples such that the slope of the solid lines is set to 1. Inset: The variation of the Toda modes in the stochastic model, for a single noise realization (for $\epsilon=10^{-3}$).}
\label{fig:mimic}
\end{figure}

The usage of projection matrices works well for small enough $\vec{dx}$, i.e., small enough kicks. For the parameters we use here, the random drive is weak enough to get a sufficiently  accurate integration of the random map in Eq.~\eqref{eq:map}.  The inset of Fig.~\ref{fig:mimic} demonstrates a single integration of our stochastic model. Clearly, $J_1$ undergoes a diffusive like motion, whereas for $k>1$ all $J_k$ do not vary in time.

We can now present the results obtained using our numerical scheme, showing how the random model gives the Lyapunov exponent of the FPUT chain. In Fig.~\ref{fig:mimic} we compare between the (quenched) average separation between trajectories in the stochastic  map and in the deterministic model. The former is obtained by simulating the random map for different pairs of  trajectories, which start close to each other and subjected the same realization of the noise $\{r_i\}$. The Lyapunov separation for the FPUT model is calculated in the standard way, using variational equations~\cite{OttBook}. We find a good agreement between the models, as is evident in the two examples presented in Fig.~\ref{fig:mimic}, for $N=63$ with $\epsilon=10^{-3}$ and $\epsilon=2.6\times 10^{-4}$. 

In all of our examples we kept the initial separation between trajectories fixed, $\Delta\vec{x}_0=10^{-8}$. One can observe in Fig.~\ref{fig:mimic} a discrepancy at short times, where the effective model shows a fast increase. This increase can be attributed to an initial ballistic separation between the trajectories~\cite{Goldfriend&Kurchan2020}. A similar, but weaker effect appears in the deterministic dynamics. The difference between the models at short times could be expected, as before the Lyapunov regime other modes than $J_1$ might affect the dynamics.  We note that the deterministic Lyapunov separation is computed in the tangent space, using variational equation, therefore, it grows indefinitely. This is opposed to the stochastic examples which are computed in phase space and saturate at long, but finite times. 

\begin{figure}
\centerline{\resizebox{0.45\textwidth}{!}{\includegraphics{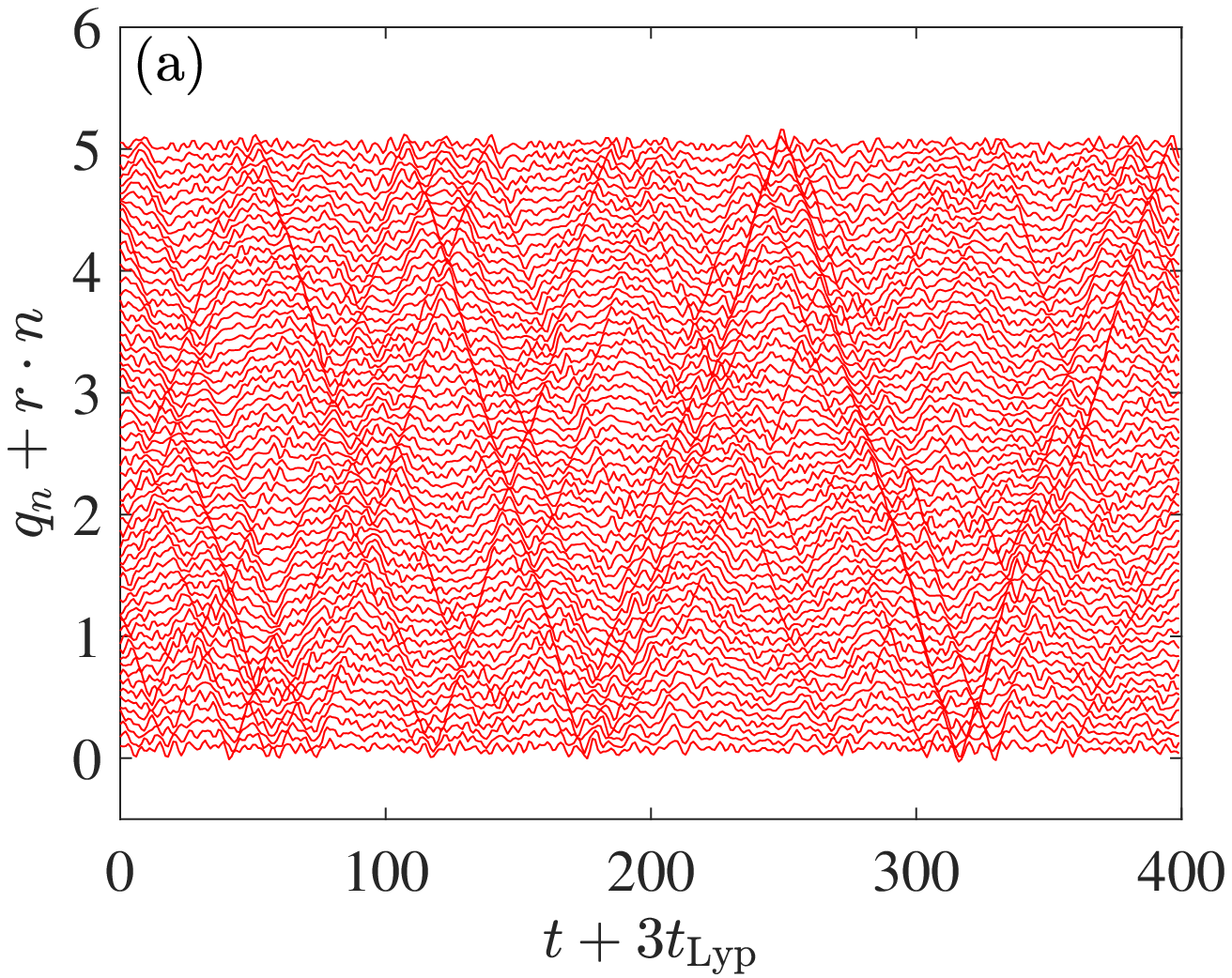}}
\hspace{1cm}
\resizebox{0.45\textwidth}{!}{\includegraphics{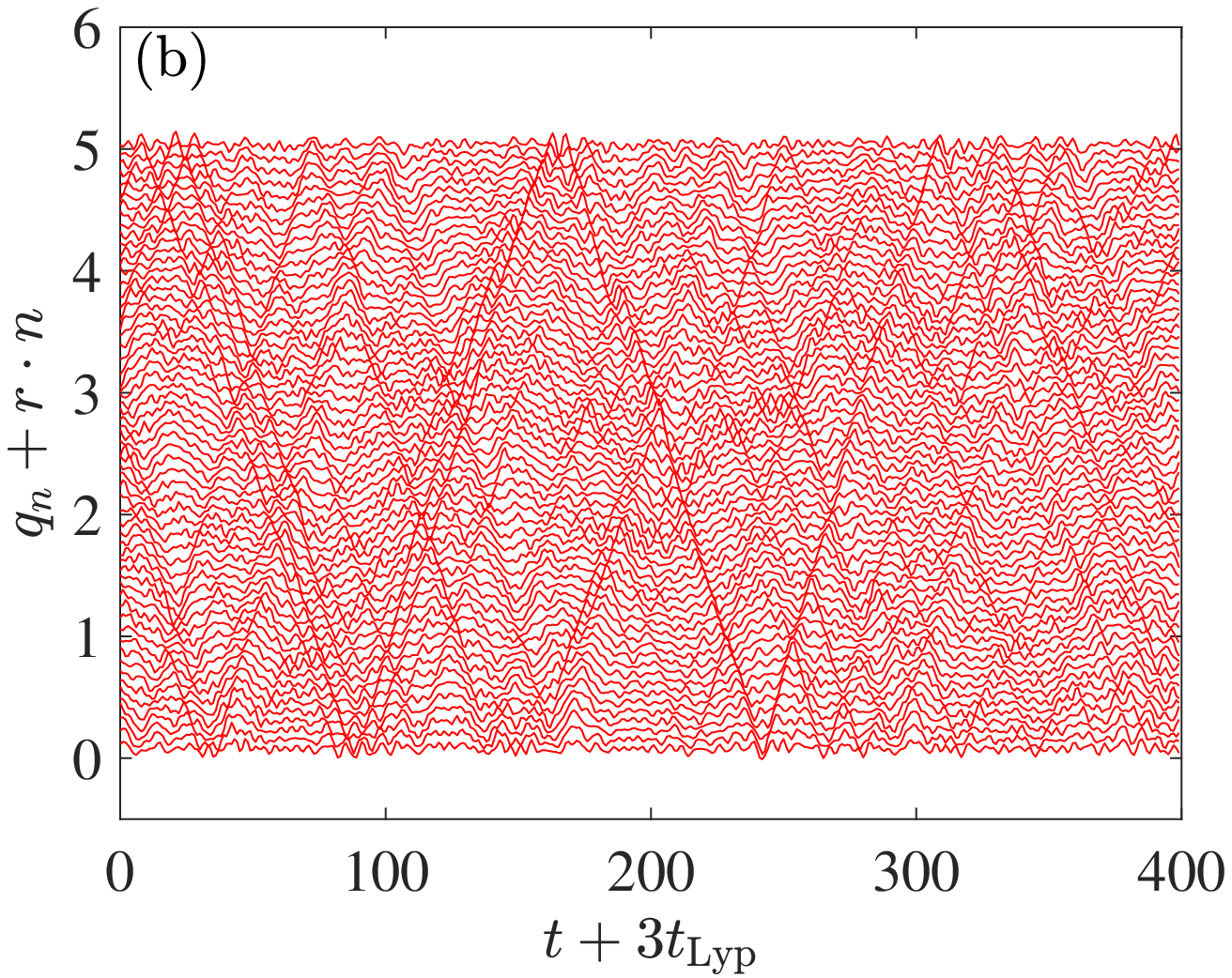}}}
\centerline{\resizebox{0.45\textwidth}{!}{\includegraphics{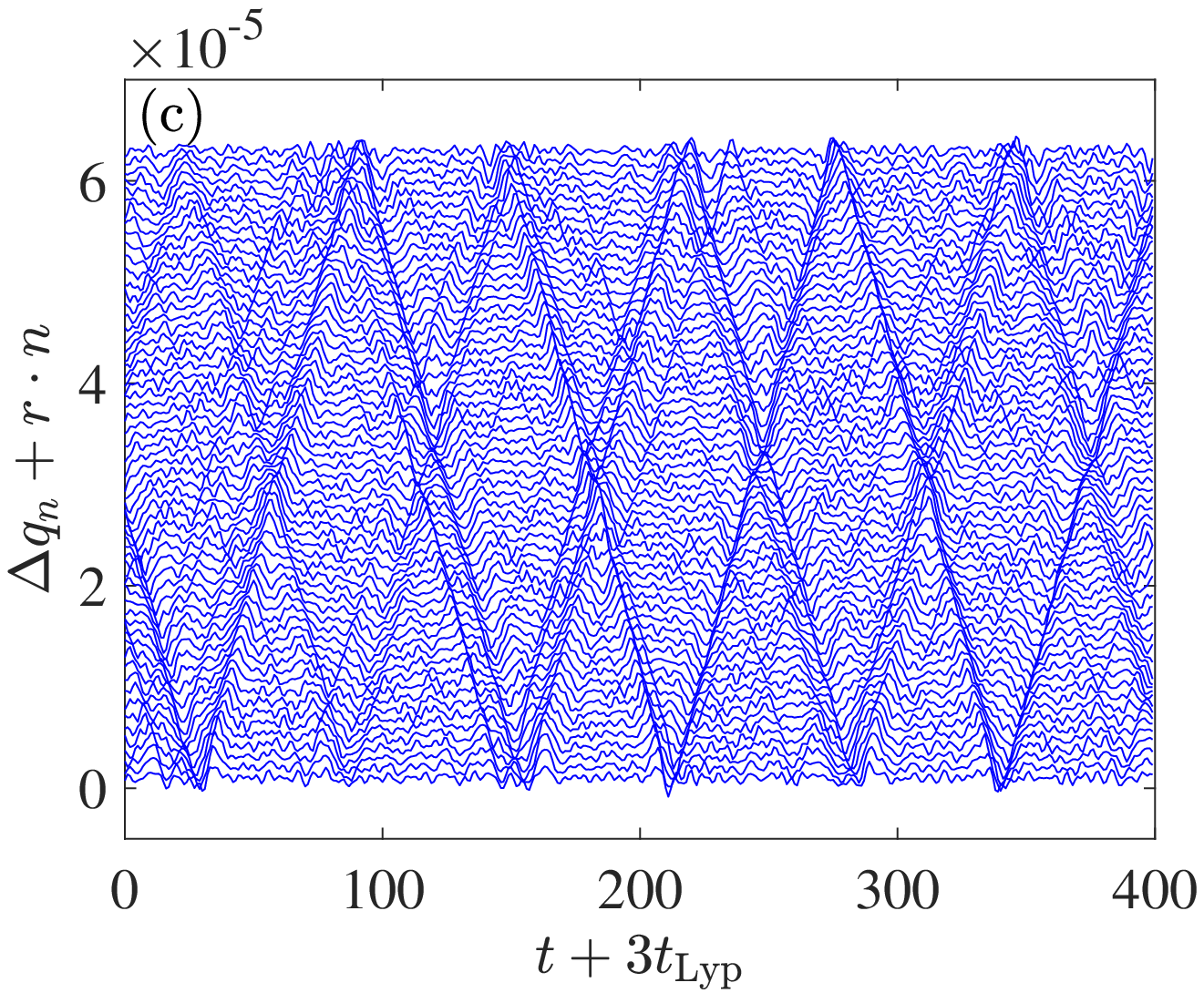}}
\hspace{1cm}
\resizebox{0.45\textwidth}{!}{\includegraphics{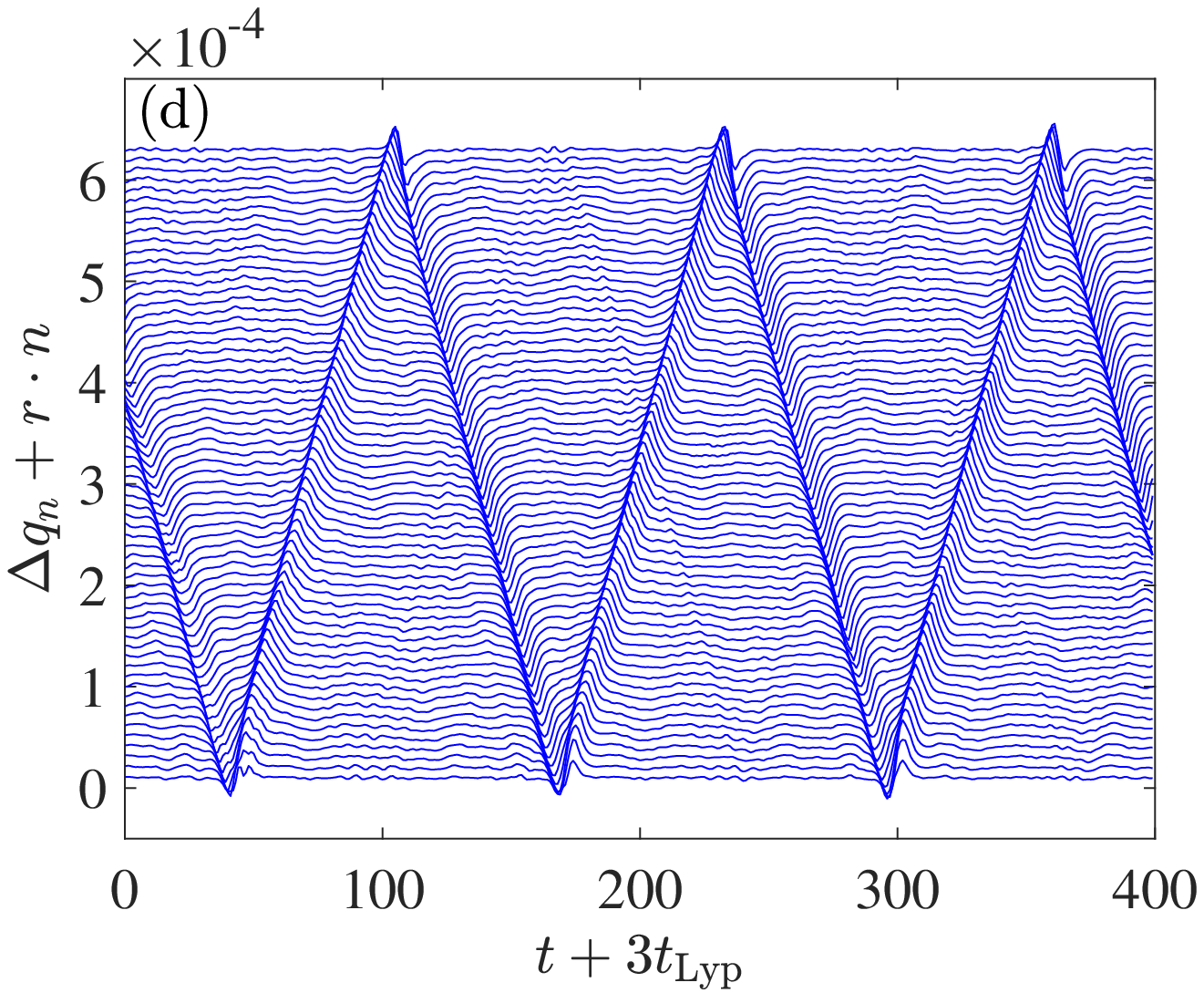}}}
\caption[]{Chain configurations as a function of time for the FPUT dynamics (panel a) and the effective model (panel b). In panels (c) and (d) we plot the projection of the  Lyapunov vector on the configuration space for the FPUT and effective model respectively. In all figures the position of the beads are equally shifted with the shift parameter $r$. The chain size is $N=63$, with an energy density of $\epsilon=10^{-3}$.}
\label{fig:solitons}
\end{figure}

Finally, we study the configuration space of the chain and the Lyapunov {\em vector} within the Lyapunov regime. In Fig.~\ref{fig:solitons} we plot the evolution of chain configurations for the two models, the deterministic FPUT and the randomly kicked Toda. In addition, we project the Lyapunov vector on the configuration space, namely, we look at $\Delta \vec{q}(t)\equiv \vec{q}^{(1)}(t)-\vec{q}^{(2)}(t)$, the difference between two trajectories which start at close proximity. All trajectories in the figure show the presence of solitons. In particular, one can clearly see that soliton excitations continue to exist in the tangent space dynamics, panels (c) and (d),  where for the effective model the Lyapunov vector concentrates along one solitary wave.

~\\~
    
\section{Conclusion and Discussion}
\label{sec:diss}

We have shown how a simple model--- a randomly perturbed Toda chain--- can explain the chaotic dynamics of the nonlinear FPUT chain at low energies:  The model provides an estimate for the largest Lyapunov exponent in quasi-Toda chains, and is consistent with the observation that the Lyapunov vector is roughly confined to the invariant Toda tori. Surprisingly, the "most integrable" element of the system, a Toda soliton mode, determines the chaotic behavior of the FPUT chain.  Let us emphasize that the random component in the stochastic model is not arbitrary; its magnitude is given by stochastic averaging of the deterministic,  multi-resonance interaction induced by all other degrees of freedom.


Integrability breaking in the FPUT chain decreases with its energy density $\epsilon$. Previous works  have shown how different stages of the near-integrable dynamics,  e.g., developing a long-lived metastable state or equipartition in the space of normal modes, happen on different time-scales, each of which vanishes as a power law $\epsilon^\nu$ (see Ref.~\cite{Benettin_etal2009} and references therein).  Resolving the origin of these power laws is crucial for understanding the phenomenon of integrability breaking in many-body systems, e.g., by identifying finite-size effects. Our model provides a simple method to obtain the behavior of $\lambda(\epsilon)$ at large quasi-Toda chains.    
We note that the Lyapunov exponent of models whose only (known) integrable approximation is the linear chain, but not the Toda model, was already addressed in previous works: For example, a scaling of $\lambda_{\beta}(\epsilon)\sim \epsilon^2$ for the $\beta$-FPUT chain, can be explained by fluctuations of curvature in the tangent space dynamics~\cite{Casetti_etal1995,Casetti_etal1996,Benettin_etal2018,Liu&He2021}.

In summary, the current Paper clearly demonstrates the idea that a weak, random perturbation to integrable models is a useful tool to treat and describe the dynamics of near-integrable systems. In addition, it highlights the different roles played by localized excitations and radiative ones, and the fact that for resolving various physical processes in the FPUT chain one must go beyond the framework of linear normal modes. In that sense, our results fit well with the understanding that "quasi-linear" and "quasi-Toda" models behave differently, e.g., as was shown recently for the energy transport in such systems~\cite{Lepri_etal2020}. Hence, our findings can stand as a building block to explore further problems in nonlinear, quasi-Toda models. Below we discuss limitations and generalizations of the stochastic model.

\subsection{Initial Conditions}

We have focused on the Lyapunov exponents for random initial conditions which are close to equilibrium. These initial conditions are drawn in the space of the normal modes of the linear chain, where the averaged energy of each normal mode $E_k$ is $\epsilon$. This choice is taken to follow previous works on Lyapunov exponents of FPUT chains, e.g., Refs.~\cite{Pettini&Landolfi1990,Casetti_etal1995,Casetti_etal1997,Benettin_etal2018} to name a few. Although we do not expect the results to change for different choice of initial conditions, this should be verified carefully. In particular, our numerical procedure for testing the effective model might be unstable for specific choice of initial conditions. If, for example, only $E_1$ is excited, then only the first few $J_k$ are excited, which means that there is a degeneracy in the spectrum of the Lax matrices $\vec{L}^{\pm}$ (recall that $J_k$ are defined via gaps between eigenvalues, see Eq.~\eqref{eq:Jk}). This can affect the numerical stability of translation in the space  of $J_k$, since it depends on finding the corresponding normal vectors $\vec{\hat{n}}_k$.  

The example mentioned above, taking initial conditions were only one, or few normal modes are excited, e.g., $E_1=\epsilon$, $E_{k\neq 1}=0$, is actually an interesting direction for future consideration. In this example not all Toda radiative modes are excited~\cite{Goldfriend&Kurchan2019}, and thus the properties of the randomly perturbed Toda model might show different behavior than the one presented in the current study. For example, the autocorrelation curves  presented in Sec.~\ref{sec:ResultA}, or the scaling of noise magnitude $\tilde{\sigma}(\epsilon)$ discussed in Sec.~\ref{sec:ResultC} can depend on the energy distribution among the Toda radiative modes.

\subsection{Thermodynamic Limit}

The maximal Lyapunov exponent of the FPUT chain depends on its size $N$~\cite{Benettin_etal2018}. It is an interesting question whether this dependence vanishes in the thermodynamic limit $N\rightarrow\infty$. The effective model presented here becomes closer to the real FPUT dynamics with increasing $N$ and decreasing $\epsilon$. This is because for larger chains we expect the integrability breaking interaction to better resemble a white noise, as coupling to more and more incoherent degrees of freedom approaches coupling to a bath. In addition, at lower energies there is a clearer separation between few soliton modes and an almost continuous spectrum of radiative modes. Hence, we claim that our approach can attack the thermodynamic limit of $\lambda(\epsilon)$.

Let us remark that the two limits of $N\rightarrow \infty$ and $\epsilon\rightarrow 0$ do not commute. For a fixed range of $\epsilon$ and increasing $N$, there is a question (posed in Ref.~\cite{Benettin_etal2018}) whether the power law exponent $\nu$, where $\lambda(\epsilon)\sim \epsilon^{\nu}$, reaches a value independent of $N$. Concerning the opposite limit, i.e., fixing $N$ and decreasing the energy $\epsilon$,  one can ask whether there is a critical value $\epsilon_c$, below which  the difference between the Toda and the linear chains is negligible~\footnote{There is an additional critical value of $\epsilon$, below which regular regions in phase-space have no zero measure. This is the KAM regime. As noted in the Introduction, this regime is expected to exist at infinitesimally small integrability breaking perturbation, and thus infinitesimally small values of $\epsilon$, which cannot be explored numerically due to the finite accuracy of the computation. }, such that the decrease of $\lambda(\epsilon)$ follows according to previous analytical estimates for quasi-linear models (e.g., for the $\beta$-FPUT chain in Refs.~\cite{Casetti_etal1995,Casetti_etal1996,Benettin_etal2018,Liu&He2021}). We have not seen any evidence of such behavior in our simulations, as well as in the results presented by Benettin et. al.~\citep{Benettin_etal2018}, which show no change in the profile of $\lambda(\epsilon)$ at very low energies.

Using our approach for large system sizes $N$ involves some computational effort, as the definition of $J_k$ depends on large matrices. However, we note that these matrices are sparse, and the quantity $J_1$ is defined only by the two largest eigenvalues. Therefore, calculating the factors that give the scaling of $\lambda(\epsilon)$ in Eq.~\eqref{eq:Lypeps} is feasible, even for very large chains $N$ and small energy densities $\epsilon$.  In more detail, the standard way to obtain the largest Lyapunov exponent involves integration of the real dynamics for $\sim 10$ Lyapunov times. On the other hand, the method presented in Sec.~\ref{sec:ResultC} allows to predict $\lambda$ at a given energy $\epsilon$ by evaluating: $\tilde{\sigma}$, $\Delta\Omega_1$, and $J_1$. Thus, the time window for which we need to integrate the system is smaller that $1$ Lyapunov time, since $\tilde{\sigma}$ corresponds to the uncorrelated apparent drive acting on $J_1$, and $\Delta\Omega_1$ refers to the integrable Toda part of the dynamics. Therefore, the only computational effort in our method is in the computation of $J_1$.

Finally, let us emphasize that we have shown how a simple equation for only the first Toda mode can capture features of the FPUT chain. While it is difficult to analyze the whole Toda spectrum, it is plausible that some analytical results can be drawn in the thermodynamic limit, in particular for $J_1$. For example, it is already known how the number density of solitons scales with the temperature of the system~\cite{Bullough_etal1993}.

\subsection{Beyond Lyapunov times}

The stochastic coarse-grained model we have treated in the current manuscript captures the dynamics of the FPUT on short time-scales, which are relevant for the chaotic separation of trajectories. However, the fact that time-dependent random perturbation can faithfully describe integrability breaking interactions in many-body systems should hold true for longer timescales. Therefore, models similar to Eq.~\eqref{eq:eff0} can explain other non-equilibrium processes in quasi-integrable systems. In particular, any process which is dictated by one or more solitons, in the apparent random background of the radiative modes, can be captured by an effective stochastic model.

One example of such process is the equilibration of the FPUT chain. In a previous work~\cite{Goldfriend&Kurchan2019} we showed how the FPUT dynamics drifts slowly between ergodized Toda tori, well characterized by $\{J_k\}$, towards equipartition. Starting with an ensemble of atypical initial conditions the average values $\langle J_k \rangle$ saturate as the system approaches  equilibration of the normal modes $E_k$, where the last Toda modes to saturate are the soliton ones. Hence, we expect that a simple model of the form $\dot{J}_1\sim v_1+g(\vec{\Omega},J_1)\eta(t)$ can capture this very last stage of equilibration. Note that now we cannot assume anymore that the multiplicative part of the noise is independent of $J_1$, as we are interested in timescales which are much larger than the Lyapunov times. In addition, the effective perturbation might be a random variable with a nonzero mean, or include a drift term $v_1$.

\begin{acknowledgments}
I would like to thank Gregory Falkovich, Jorge Kurchan, Alberto Maspero, and Yuval Peled for useful suggestions and discussions. The work was supported by the Scientific Excellence Center at WIS, grant  662962 of the Simons  foundation, grant 075-15-2019-1893 by the Russian Ministry of Science,  grant 873028 of the EU Horizon 2020 programme, and grants of ISF and BSF.
\end{acknowledgments}

\appendix

\section{Adding a drift term}
\label{sec:drift}

In Sec.~\ref{sec:ResultB} we mention that for the numerical integration of the stochastic model it might be useful to add a drift term to the dynamics:
\begin{equation}
\dot{J}_1 = v_1(\vec{J}) + f(\vec{J},\vec{\Psi})\eta(t).
\end{equation} 
The idea is to choose $v_1$ which is small enough such that it does not effect the Lyapunov exponent, yet, large enough to  guarantee that $J_1$ does not approach zero during the simulation (recall that $J_1\geq 0$).   We now discuss this point in detail.

For simplicity we consider the one-dimensional problem in Eq.~\eqref{eq:effgen} with an additional drift term for the action variable $I$. The tangent space equations dictating the Lyapunov separation is given by 
\begin{align}
\begin{split}
\dot{dI} &= \frac{\partial v(I) }{\partial I}dI+ \left(\frac{\partial g(I,\theta)}{\partial  \theta}d\theta+\frac{\partial g(I,\theta)}{\partial  I}dI\right)\eta(t), \\
\dot{\theta} &=\frac{\omega(I)}{\partial I}dI.
\end{split}
\label{eq:tanIth}
\end{align}
Under the assumption that $I$ is constant we can eliminate the drift term by moving to a new variable $\tilde{dI}=dIe^{-\frac{\partial v(I) }{\partial I}t}$. Thus, the Lyapunov exponent of the system in Eq.~\eqref{eq:tanIth} is given by adding $\frac{\partial v_1(J_1) }{\partial J_1}$ to the expression in Eq.~\eqref{eq:Lyp}. Therefore, a drift term which satisfies $v_1(J_1)\lambda^{-1}\ll J_1$ and  $\frac{\partial v_1(J_1) }{\partial J_1} \ll \lambda$ does not effect the Lyapunov exponent of the effective model without a drift in Eq.~\eqref{eq:eff0}. 

For the numerical simulations in Sec.~\ref{sec:ResultB} we add a drift term to $J_1$, choosing $v=a/J_1$ with $a\sim \tilde{\sigma}$, which guarantees that $J_1$ does not approach zero during the simulation. We have verified that the Lyapunov exponent, as well as the typical variation in $J_1$, are not affected by $v$.

\section{Verifying the numerical methods}
\label{sec:Num}

The different numerical methods applied in the current work include different computational errors, and extra care shall be taken when dealing with the analysis of quantities which are, theoretically, conserved by the dynamics. We now discuss several aspects that were tested for the numerical simulations. 

Let us first treat the most natural question: We try to evaluate the small breaking of integrability in the FPUT dynamics, characterized by the variance $\tilde{\sigma}$;  Can it be that the measured values shown in Fig.~\ref{fig:scaling}(a) simply originate from the inaccuracy of the numerical integration, and not from the FPUT dynamics? Indeed the values of $\tilde{\sigma}$ presented in that figure are rather small.  To answer this question  we repeat the same procedure for obtaining $\tilde{\sigma}(\epsilon)$, as outlined in Sec.~\ref{sec:ResultC}, but for the {\it integrable Toda} evolution. The values found for $\tilde{\sigma}_{\rm Toda}(\epsilon)$ are of order $10^{-22}$, much smaller than the ones presented in Fig.~\ref{fig:scaling}(a). The only place where the  "numerical error breaking of integrability" found to be comparable to the one attributed to the FPUT is when we treat the $\gamma_T$ model at very low energies of $\epsilon\sim 10^{-5}$.

Apart from the essential check for the computation of $\tilde{\sigma}$, we have verified all the other elements of the numerical analysis: (I)~As already mentioned above, we have checked that the symplectic integration conserves energy and the Toda modes with a sufficient accuracy. (II)~The calculation of $J_k$ is based on finding eigenvalues of big, although sparse, matrices. Numerical errors can affect the computations of $\dot{J}_k=(J_k(t)-J_k(t+h))/h$, due to the FPUT evolution in small time-interval $h$. We have tested different values of $h$ to verify that our choice of $h\sim 10^{-4}-10^{-6}$ is suitable. (III)~The motion in the space of Toda modes, e.g., kicking the system in the $J_1$ direction alone, involves computation of eigenvectors (which gives $\hat{\vec{n}}_k$ and projection matrices). It is hard to evaluate the error of this operation, as it depends on many parameters such as the magnitude of the kick itself. However, this computational issue is irrelevant for our results: whenever the changes in $J_{k>1}$ are 1-2 orders of magnitude smaller than changes in $J_1$, which undergoes a random motion, we have found a Lyapunov growth rate similar to the one of the deterministic FPUT.

Let us remark that we did encounter several examples were we could not kick the system along $J_1$ alone. This instability of the algorithm is not due to the magnitude of the kicks, but comes from the projection matrices themselves, and thus, it is  probably related to singularities in the calculation of $\vec{\hat{n}}_k$.

\bibliography{PerturbedToda}

\end{document}